\documentclass[sigconf]{acmart}
\usepackage{algorithm}
\usepackage{graphicx}
\usepackage{multirow}
\usepackage{booktabs}
\usepackage[tight,footnotesize]{subfigure}
\usepackage{algorithmic}
\usepackage{amsfonts}
\usepackage{amsmath}

\AtBeginDocument{%
  \providecommand\BibTeX{{%
      \normalfont B\kern-0.5em{\scshape i\kern-0.25em b}\kern-0.8em\TeX}}}

\copyrightyear{2024} \acmYear{2024} \setcopyright{rightsretained}
\acmConference[HotCarbon'24]{The 3rd ACM HotCarbon Workshop on Sustainable
  Computer System}{9 July 2024}{Santa Cruz, CA} \acmBooktitle{The 3rd ACM
  HotCarbon Workshop on Sustainable Computer System (HotCarbon '24), 9 July
  2024, Santa Cruz, CA}
\begin{document}

\title{Offline Energy-Optimal LLM Serving: Workload-Based Energy Models for LLM
  Inference on Heterogeneous Systems}

\author{Grant Wilkins}
\email{gfw27@cam.ac.uk}
\affiliation{%
  \institution{University of Cambridge}
  \city{Cambridge}
  \country{UK}
}


\author{Srinivasan Keshav}
\email{sk818@cam.ac.uk}
\affiliation{%
  \institution{University of Cambridge}
  \city{Cambridge}
  \country{UK}
}

\author{Richard Mortier}
\email{rmm1002@cam.ac.uk}
\affiliation{%
  \institution{University of Cambridge}
  \city{Cambridge}
  \country{UK}
}

\begin{abstract}
  The rapid adoption of large language models (LLMs) has led to significant
  advances in natural language processing and text generation. However, the
  energy consumed through LLM model inference remains a major challenge for
  sustainable AI deployment. To address this problem, we model the
  workload-dependent energy consumption and runtime of LLM inference tasks on
  heterogeneous GPU-CPU systems. By conducting an extensive characterization
  study of several state-of-the-art LLMs and analyzing their energy and runtime
  behavior across different magnitudes of input prompts and output text, we
  develop accurate $(R^2>0.96)$ energy and runtime models for each LLM. We
  employ these models to explore an offline, energy-optimal LLM workload
  scheduling framework. Through a case study, we demonstrate the advantages of
  energy and accuracy aware scheduling compared to existing best practices.
\end{abstract}



\maketitle

\section{Introduction}
Rapid advancements in large language models (LLMs) have revolutionized natural
language processing, enabling AI systems to achieve human-level performance on a
wide range of language tasks~\cite{openai2023gpt4, bommasani2022opportunities,
  geminiteam2024gemini}. However, the computational resources and energy
consumption associated with deploying these models present significant
challenges to not only energy systems but also sustainability
goals~\cite{patterson2021carbon,lin2024exploding,lin2023adapting}. As LLMs
become increasingly integrated into real-world applications, optimizing their
energy efficiency during inference is crucial for sustainable AI
development~\cite{wu2022sustainable}.

Inference, the process of using a trained model to make predictions on new data,
is a critical phase in LLM deployment as it is the point at which AI
capabilities become accessible to users. Unlike the one-time training process,
inference is an ongoing, real-time process that directly impacts end-user
experience. Inference in LLMs can be computationally expensive due to model
size~\cite{wu2022sustainable} and quality of service/latency
expectations~\cite{wang2024efficient}. Scaling LLMs up across large data centers
is challenging due to power~\cite{patel2024characterizing} and communication
overheads~\cite{patel2023splitwise}.

The energy intensity of LLM inference can be substantial even when compared to
training~\cite{chien2023reducing}. Decarbonizing the energy sources for
data centers can be challenging due to both sporadic demand and regional
inefficiencies in adopting renewables. Higher energy consumption of an
application approximately correlates with greater carbon
intensity~\cite{radovanovic2022carbon}. It is thus crucial to find
energy-efficient methods to mitigate the environmental costs of LLM
inference~\cite{anderson2023treehouse, li2024environmentally}.

To address this issue, we propose a workload-based model of energy
consumption for LLM inference to let system operators navigate the trade-off
between accuracy and energy usage.

Our contributions are as follows:

\begin{enumerate}
\item We characterize the energy consumption and runtime behavior of several
  state-of-the-art LLMs on a heterogeneous GPU-CPU system~(\S\ref{sec:routing}).
\item We develop workload-based energy and runtime models that accurately
  capture the relationship between the number of input and output tokens and the
  energy and runtime characteristics of each LLM~(\S\ref{sec:results}).
\item We demonstrate the effectiveness of our approach through a case study,
  showcasing a tunable trade-off between energy and
  accuracy~(\S\ref{sec:fitting}).
\end{enumerate}
Our profiling framework and datasets are openly
available.\footnote{\url{https://github.com/grantwilkins/energy-inference.git}}



\section{Related Work}\label{sec:related-works}

\subsection{Energy Consumption in AI Systems}
Recent reports have found that the computation required by state-of-the-art AI
systems entail massive energy consumption and carbon
emissions~\cite{samsi2023words, chien2023reducing, wu2022sustainable,
  luccioni2022estimating, patterson2021carbon}. The energy intensity of AI
systems can be broadly split between the energy required for training and that
required for inference after models are deployed~\cite{henderson2020towards}.
Training complex models on massive datasets is an energy-intensive process, with
estimates finding that training GPT-3 required 1,287 megawatt-hours of
energy~\cite{patterson2021carbon}. Even with this huge amount of energy, a year
of inference by an LLM on cloud infrastructure can consume over 25$\times$ more
energy than training that same model~\cite{chien2023reducing}. Some of these
issues and emissions of course depend on the deployment scale and hardware
efficiency~\cite{samsi2023words}, however, the trend remains that energy
consumption in inference is a large issue. Optimizing software and hardware
specifically for AI workloads is thus essential~\cite{anderson2023treehouse}.

Desislavov et al.~\cite{radosvet2023trends} provide an examination of trends in
AI inference energy consumption, arguing that while performance has increased
dramatically, energy consumption has not escalated at the same pace, thanks to
hardware optimizations and algorithmic innovations. Chien et
al.~\cite{chien2023reducing} discuss larger trends in LLM inference energy
consumption and do not focus on device-level energy modeling benefits. Samsi et
al.~\cite{samsi2023words} explore the energy consumption of Meta's Llama LLMs
for different batch sizes and numbers of GPUs, showing the potential energy
reductions obtainable by tuning these parameters. Stojcovik et
al.~\cite{stojkovic2024greener} discuss the impacts of GPU frequency scaling on
the energy efficiency of serving LLMs; however, at this point, this work is only
a characterization and not an applied analysis.

Our work extends these studies with a thorough CPU+GPU energy measurements
across multiple model families and sizes, producing one of the most
comprehensive datasets of its kind.

\subsection{Energy-Aware Data Center Scheduling}
A large body of work that focuses on energy-aware
scheduling~\cite{tang2020cpu,ramapantulu2014modeling, mei2017energy,
  huo2012energy, fan2023synergy, liang2023model}, but none of these focus on the
unique challenge of developing workload-aware models for LLM inference towards
this goal. Hu et al.~\cite{helios2021hu} analyze deep learning workloads in GPU
data centers, offering insights into energy conservation strategies through
workload scheduling. This research aligns with our objectives by confirming the
critical role of scheduling in reducing energy footprints.

Li et al.~\cite{li2023clover} introduce Clover, which promises to minimize
carbon emissions for serving AI inference. Unlike our study, this work does not
explicitly consider LLMs or a per-model function to capture energy and runtime,
instead focusing on carbon-emission patterns for a data center.

Gu et al.~\cite{gu2023powerflow} presents PowerFlow, a tool that uses
clock-frequency data from GPUs to minimize energy consumption as a scheduling
decision. However, their study does not consider LLMs and is not necessarily
workload-aware.

Patel et al. introduce POLCA~\cite{patel2024characterizing}, which can provide a
way to automatically power-cap based on existing workload traces. Li et
al.~\cite{li2024environmentally} focuses on delivering a geographic load
balancing perspective for AI inference, optimizing environmental equity.
However, their model considers large-scale workload traces, not device-level
energy and runtime data.

Our work aims to fill the niche with energy-aware LLM inference scheduling using
measurements from state-of-the-art open-source LLMs leading to an applied
analysis using offline optimization. The results of our findings can be used by
system operators to accurately predict and schedule based on the amount of
energy and runtime for inference.

\section{Methods}\label{sec:methods}
For our LLM inference engine we use Hugging Face's Accelerate~\cite{accelerate}.
This library uses all available GPUs, and divides a model among the available
GPUs in a tensor parallel fashion to minimize intermediate communication and
maximize the distributed capabilities for computation across the devices. We
disable KV-caching~\cite{attention} to ensure that our measurements are
consistent between runs and do not require a warm-start phase.




\subsection{LLM Selection}\label{subsec:models}

We study several open-source LLMs, summarized in Table~\ref{tab:models}. By
profiling different LLMs we are able to explore the effects of diverse
architectures and parameter values on runtime, energy consumption, and accuracy.
For each model, we conduct a series of standardized text generation prompts to
evaluate their energy consumption during inference.

Numerous benchmarks have sought to quantify LLM accuracy, e.g.,~the
MMLU~\cite{hendrycks2021measuring} and HellaSwag~\cite{zellers2019hellaswag}. To
avoid the inadequacies introduced by individual tests for
accuracy~\cite{mcintosh2024inadequacies}, we use the Hugging Face
Leaderboard's~\cite{open-llm-leaderboard} \textit{average accuracy}, denoted
$A_K,$ that averages the performance of a model, $K,$ on a large repository of
datasets and tests.

\begin{table}[!htb]
  \centering
  \caption{LLM Energy Consumption and Runtime}
  \label{tab:models}
  \begin{tabular}{lrrr}
    \toprule
    \textbf{LLM (\# Params)} & \textbf{vRAM Size (GB)} & \textbf{\# A100s} & \boldmath$A_K(\%)$~\cite{open-llm-leaderboard} \\
    \midrule
    Falcon (7B)     &  14.48 & 1 & 44.17 \\
    Falcon (40B)    &  83.66 & 3 & 58.07 \\
    Llama-2 (7B)    &  13.48 & 1 & 50.97 \\
    Llama-2 (13B)   &  26.03 & 1 & 55.69 \\
    Llama-2 (70B)   & 137.98 & 4 & 64.52 \\
    Mistral (7B)    &  15.00 & 1 & 60.97 \\
    Mixtral (8x7B)  &  93.37 & 3 & 68.47 \\
    \bottomrule
  \end{tabular}
\end{table}

\subsection{Energy Profiling of Our Cluster}\label{sec:systems}
We perform all experiments the Swing cluster at Argonne National Lab using a
single node with $8\times$Nvidia A100 (40GB) GPUs, $2\times$AMD Epyc 7742
64-core processors, and 1TB of DDR4 RAM. We use only the minimum number of GPUs
as shown in Table~\ref{tab:models}. We profile the system's energy consumption
during inference using tools that capture Nvidia GPU energy and AMD CPU power
while timing the operation. Our methods utilize the known relationship that
$E = Pt$ where $E$ represents energy, $P$ is average power, and $t$ is runtime.

\subsubsection{NVIDIA GPUs}
We use PyJoules~\cite{pyJoules}, a Python-based energy measurement library, to
quantify the energy consumption associated with inference on NVIDIA GPUs.
PyJoules provides an interface to \texttt{NVML}~\cite{nvidia_nvml}, providing a
software-defined energy usage assessment for targeted NVIDIA devices. This tool
offers GPUs real-time energy consumption for a given tracked process.

\subsubsection{AMD CPUs}
We adopt a different strategy for AMD CPUs due to the absence of a Python API.
Instead, we utilize AMD$\mu$Prof's \texttt{timechart} feature, which provides
detailed power draw metrics for every core on the chip at fine-grained
intervals. By polling AMD$\mu$Prof at 100ms intervals, we can capture the power
draw of each physical CPU core throughout the model inference process.

To ensure we accurately attribute the energy consumption to our inference task,
we monitor the CPU core residency through \texttt{psutil}. This information
allows us to identify and record the specific cores actively engaged in the
inference process at each time step. The total energy consumption for the
inference task is then calculated by summing the power usage across all active
cores and summing over the product of the power usage and time of inference, as
follows:
\begin{equation*}
  E_{Total, CPU} = \sum_{core}{\left(\sum_{i}{P_{core, i} \Delta t_i}\right) }
\end{equation*}
where $P_{core, i}$ represents the power draw of an individual core at each time
step, $i,$ with $\Delta t_i$ being the time step size.

\section{Problem Formulation}\label{sec:routing}
The purpose of developing workload-based models of LLM inference is to create a
framework that allows a data center operator to navigate the trade-off between
model accuracy and energy consumption. To do so, we formalize an optimization
problem below.

Consider a data center that hosts $\mathcal{K}=\{1, \dots, K\}$ distinct LLM
models. Assume that a fraction $\gamma_K$ of the inference workload is assigned
to each model $K$, where $\gamma_K \in [0, 1], \forall K$ and
$\sum_{K \in \mathcal{K}} \gamma_K = 1.$

We denote a query $q$ by its count of input and output tokens,
$q = (\tau_{in}, \tau_{out})$. A workload with $m$ queries is then a multiset
$Q \in \left(\mathbb{N}^2\right)^m$. As our goal is to perform scheduling of
each query, we must create a disjoint partition of our set $Q$. We say that each
$Q_K \in \left(\mathbb{N}^2\right)^{m_K}$ has $m_K$ prompts and is composed of a
set of lengths of input and output tokens
$Q_K = \left\{ (\tau_{in, 1}, \tau_{out, 1}), \dots, (\tau_{in, i}, \tau_{out,
    i}) \right\}.$

Since this is an offline setting we assume we have perfect knowledge of our
system, including the number of output tokens that a given input prompt will
produce. In reality, this is not known \textit{ab initio} though work by Zheng
et al.~\cite{zheng2023response} has shown that the number of output tokens can
be reasonably well estimated by analyzing past input-output pairs.

For optimization purposes, we must define a function based on the constant $A_K$
from Table~\ref{tab:models}. We propose $a_K: \mathbb{N}^2 \to [0, \infty),$ a
monotonically increasing function based on the number of input and output tokens
that a model $K$ ingests and produces. Therefore, for a model $K$ processing
tokens $(\tau_{in}, \tau_{out})$ we have
\begin{equation}\label{eqn:accuracy}
  a_K(\tau_{in}, \tau_{out}) = A_K \tau_{in} + A_K\tau_{out}.
\end{equation}
As we will later derive, we denote a model for energy consumption for a given
number of input and output tokens as
$e_{K}(\tau_{in, i}, \tau_{out, i}): \mathbb{N}^2 \to [0, \infty)$.

Both of these functions have a normalized counterpart
$\widehat{e_{K}}, \widehat{a_{K}}: \mathbb{N}^2 \to [0,1]$ that scales the cost
associated with these values $[0,1]$ to make these different metrics comparable.
We normalize by dividing by the largest known value of energy and accuracy prior
to optimization.

Finally, let $\zeta \in [0,1]$ denote a tuning parameter that lets a data center
operator trade off energy for accuracy. Let $|Q|$ represent the total number of
queries in our workload, and $|Q_K|$ represent the total number of queries each
model $K$ processes.

We now formulate our workload assignment problem as:
\begin{align}
  \underset{{Q_K \in Q}}{\min} \
  &\sum_{K \in \mathcal{K}}{\sum_{(\tau_{in}, \tau_{out}) \in Q_K} \zeta \widehat{e_K}(\tau_{in}, \tau_{out})  - (1-\zeta)\widehat{a_K}(\tau_{in}, \tau_{out})} \label{eqn:opt-2-1} \\
  &\text{s.t.} , 0 < \frac{|Q_K|}{|Q|} < 1\label{eqn:opt-2-2}\\
  &Q = \bigcup_{K \in \mathcal{K}} Q_K\label{eqn:opt-2-3} \\
  &Q_I \cap Q_J = \emptyset, I \neq J, \forall I, J \in \mathcal{K},\label{eqn:opt-2-4}
\end{align}
where Equations \ref{eqn:opt-2-3} and \ref{eqn:opt-2-4} define the partition
coverage of the workload, and Equation \ref{eqn:opt-2-2} ensures we give each
LLM some queries. In our implementation, we dynamically normalize our energy and
accuracy measures across all the queries to allow us to adjust the scale of
costs across different models and query combinations. This problem is
computationally intensive to solve as it is an example of a general assignment
problem which are known to be NP-hard~\cite{fisher1986multiplier}.

\section{LLM Inference Performance}\label{sec:results} 
All hardware information we state in Section~\ref{sec:systems}. We use Ubuntu
20.04 with Python 3.12.0, PyTorch v2.0.1, Torchvision v0.15.2, Numpy v1.26.0,
Hugging Face v0.20.2, and Accelerate v0.26.1.

\subsection{Experimental Strategy}

We conduct an experimental campaign to evaluate the performance of differing
workloads across various models. We systematically varied the number of input
and output tokens to measure their effects on runtime and energy consumption
under two main experimental conditions. In each experiment we do not allow for
key-value caches~\cite{attention} to be re-used to ensure our measurements are
standard between iterations. We fix the batch size at 32.

\subsubsection{Vary Input Tokens} For the first experimental condition, we
executed inference requests with increasing the number of input tokens, ranging
from 8 to 2048 tokens, while maintaining a fixed output token size of 32. This
setup allowed us to isolate the impact of input size on the system's performance
and energy efficiency.

\subsubsection{Vary Output Tokens} In the second set of experiments, we varied
the output token limit from 8 to 4096 tokens, keeping the number of input tokens
constant at 32. This approach helped us understand how increasing output demands
affect the runtime and energy consumption of the systems tested.

\subsubsection{Randomization and Stopping Criteria}
Each experiment was conducted in a randomized order to mitigate any potential
bias introduced by the sequence of tests. Also, we repeated trials until either
of two conditions was met: (\emph{i})~the measured runtime was within 0.5
seconds of the actual mean runtime with 95\% confidence; and (\emph{ii})~a
maximum of 25 trials were conducted for each setting if the first condition
could not be met.

\subsection{Input Token Effects}

\begin{figure*}[!htb]
  \centering
  \captionsetup{justification=centering}
  \subfigure[Runtime]{
    \label{fig:input-runtime}
    {\includegraphics[height=32ex]{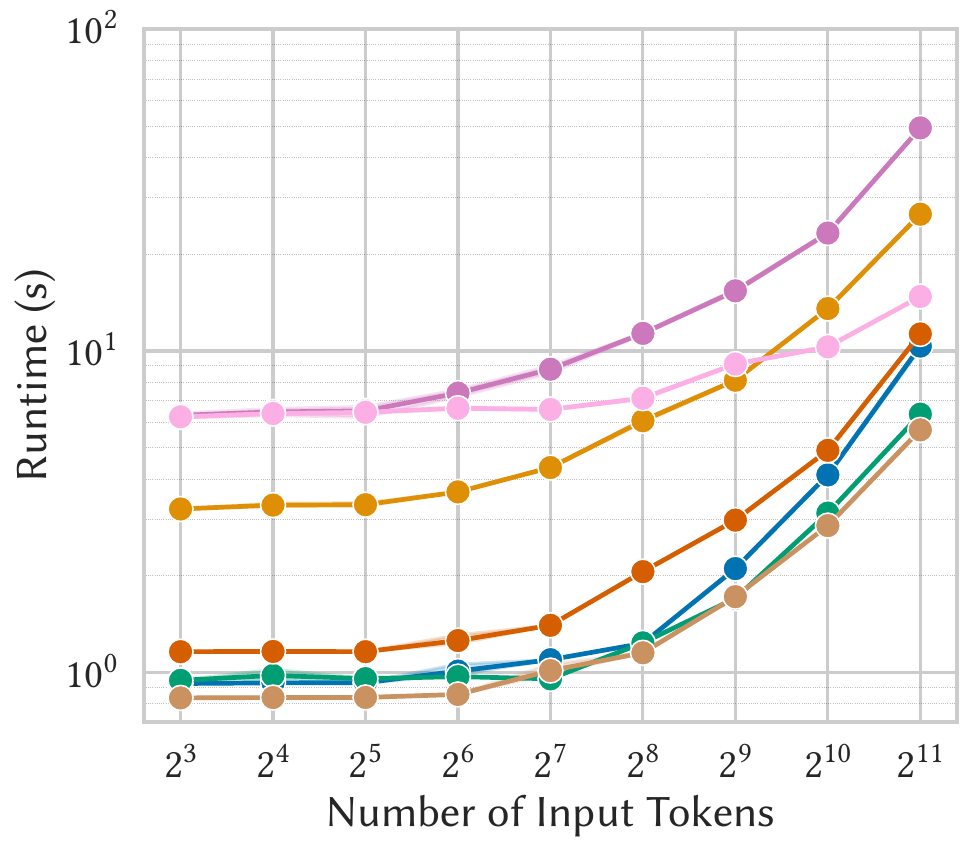}}
  }
  \subfigure[Throughput]{
    \label{fig:input-throughput}
    {\includegraphics[height=32ex]{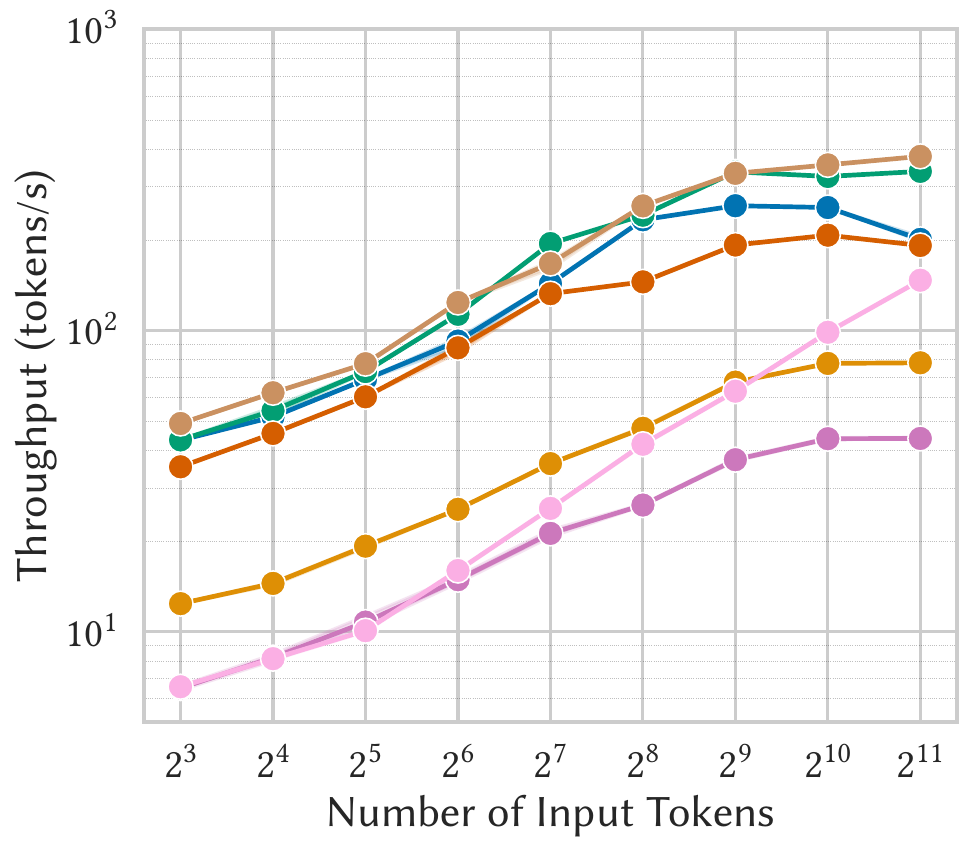}}
  }
  \subfigure[Energy per Token]{
    \label{fig:input-energy-per-token}
    {\includegraphics[height=32ex]{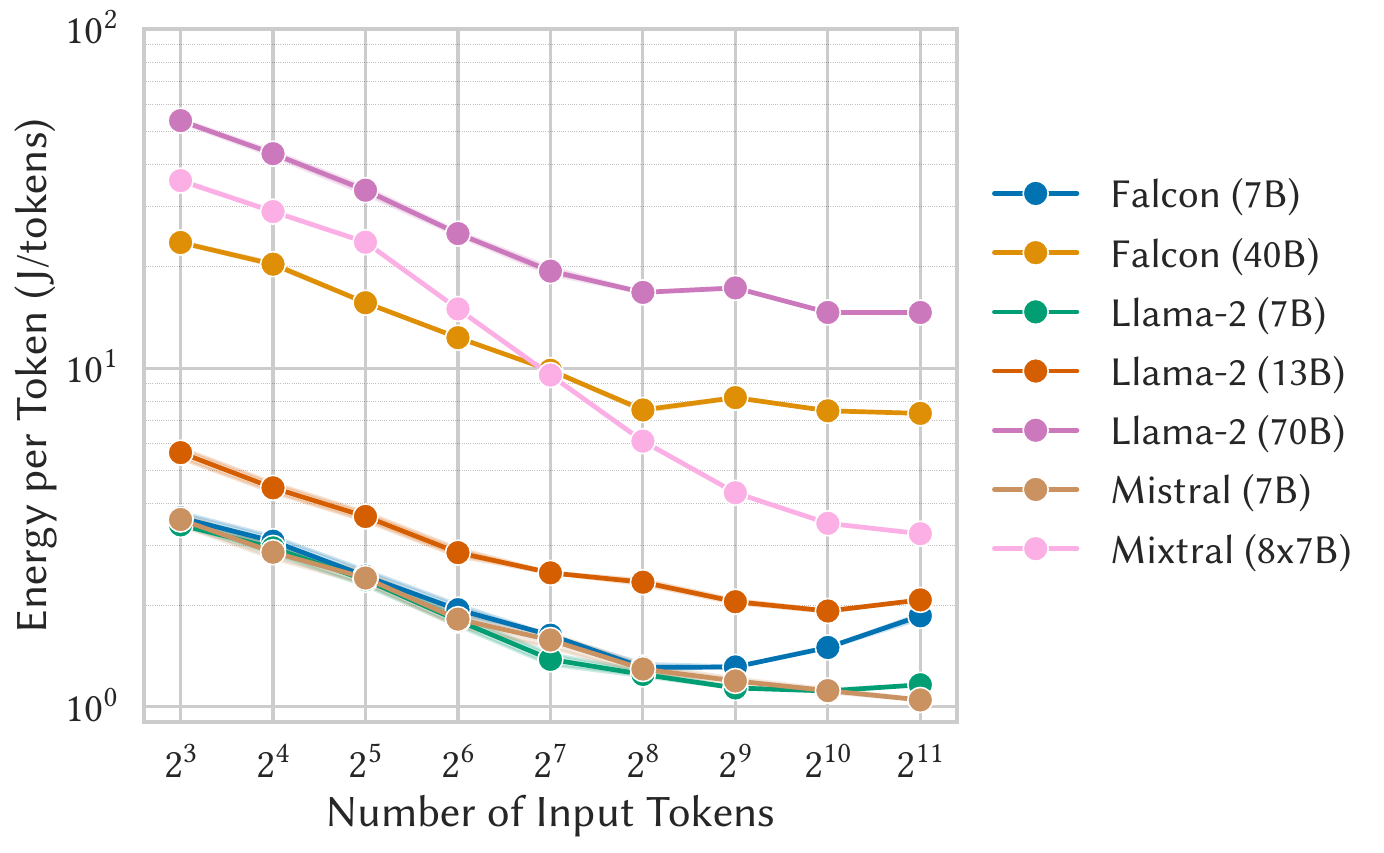}}
  }
  \caption{\label{fig:input-token-results}Model performance against number of
    input tokens. Low variance renders error bars invisible.}
\end{figure*}

\begin{figure*}[!htb]
  \centering
  \captionsetup{justification=centering}
  \subfigure[Runtime]{
    \label{fig:output-runtime}
    {\includegraphics[height=31ex]{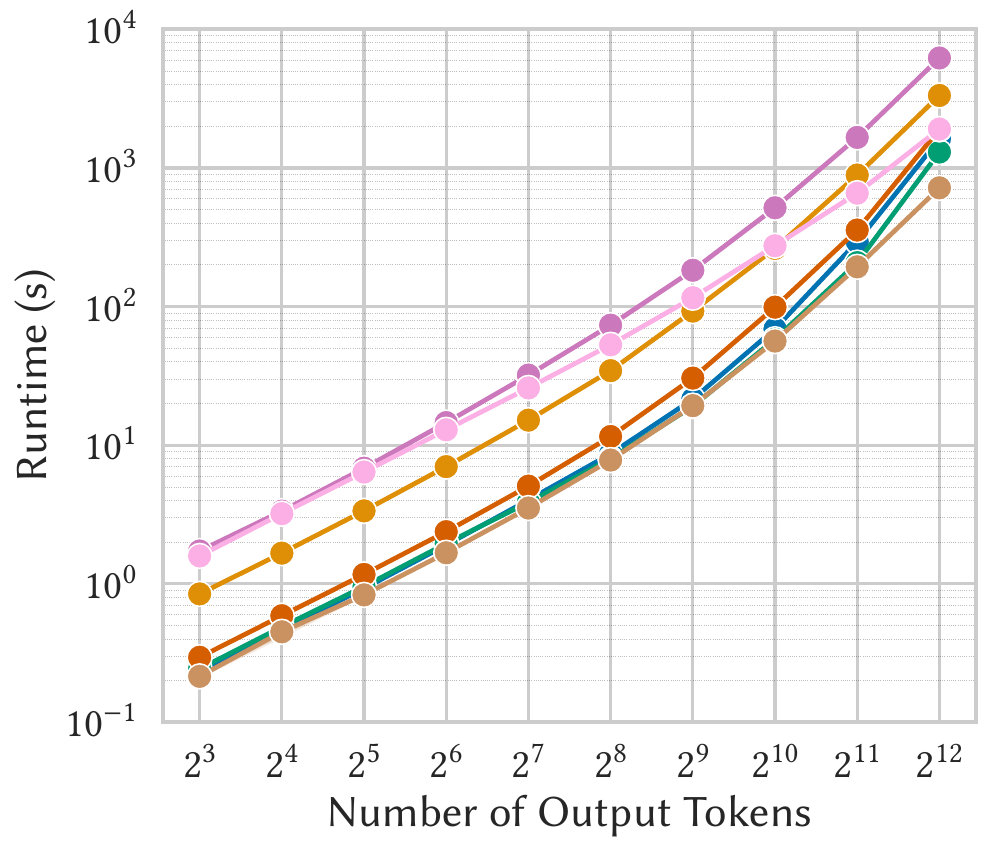}}
  }
  \subfigure[Throughput]{
    \label{fig:output-throughput}
    {\includegraphics[height=31ex]{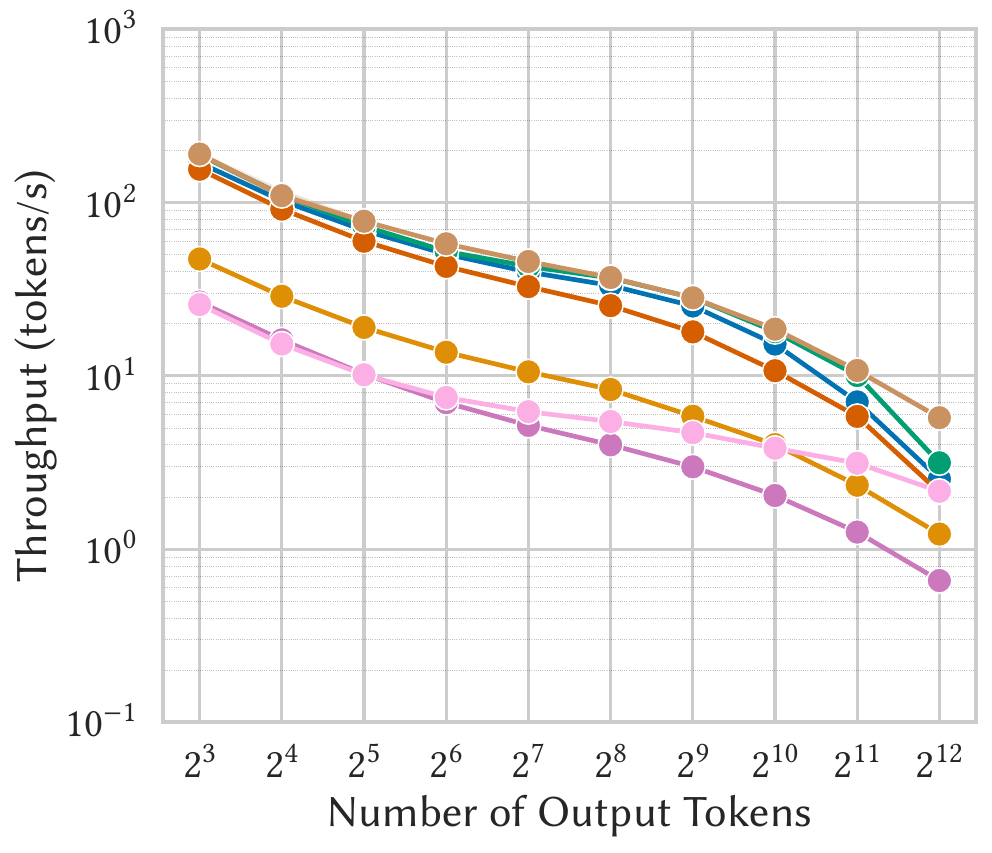}}
  }
  \subfigure[{Energy per Token}]{
    \label{fig:output-energy-per-token}
    {\includegraphics[height=31ex]{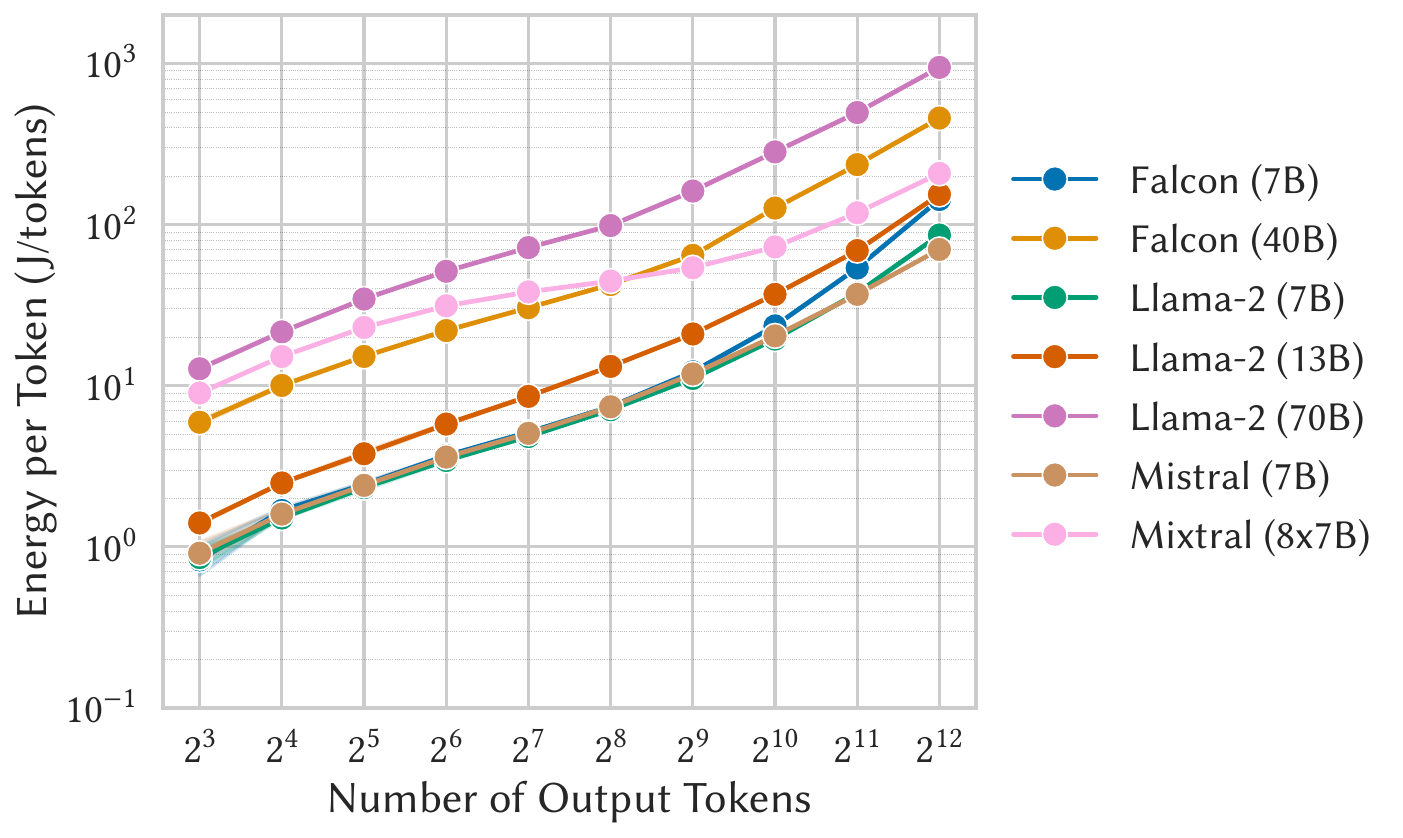}}
  }
  \caption{\label{fig:output-token-results}Model performance against number of
    output tokens. Low variance renders error bars invisible.}
\end{figure*}

Figure~\ref{fig:input-token-results} presents the impact of varying numbers of
input tokens on the runtime, throughput, and energy per token for various LLMs.
The results depict a clear trend: as the number of input tokens increases, the
runtime tends to increase, while the throughput plateaus, in accordance with a
roofline model~\cite{roofline}. Specifically, the runtime increase is most
pronounced for larger models like Llama-2 (70B) and Falcon (40B), likely due to
the higher computational burden these models sustain as they process more
extensive input sequences. The energy consumption per token demonstrates similar
trends, with smaller models exhibiting lower energy per token compared to larger
models.

An outlier to all of these cases is Mixtral (8x7B), which has a higher
throughput and energy efficiency compared to other large models at larger token
input sizes. This LLM's sparse mixture-of-experts architecture
(SMoE)~\cite{jiang2024mixtral, pmlr-v162-rajbhandari22a} allows it to activate
just 12B parameters on average by selecting two expert sub-models. This
classification phase comes with an added runtime and energy overhead, however,
on larger prompts it regains its performance capabilities. Therefore, for SMoE
one gets the accuracy advantages of a large model for less energy and lower
runtime than its denser counterparts.

\subsection{Output Token Effects}
Figure~\ref{fig:output-token-results} illustrates how changes in the number of
output tokens affect runtime, throughput, and energy consumption per token
across different LLMs. Notably, the runtime exhibits a steep increase with
larger output token sizes, which is consistent across all models but is
especially significant for the high-parameter models such as Falcon (40B) and
Llama-2 (70B). Throughput, decreases as the number of output tokens increases.
This inverse relationship highlights the additional time required to generate
each additional token, which involves more extensive interaction between model
layers and successive passes through the LLM to generate each
token~\cite{attention}. Energy per token also increases with the number of
output tokens and number of parameters. This increase is particularly sharp in
higher-parameter models like Falcon (40B).

Again, Mixtral (8x7B) demonstrates greater energy efficiency compared to its
large parameter counterparts. Even in cases of high output token generation, an
SMoE architecture can yield improvements in energy efficiency.

\section{Workload-Based Model Fitting}\label{sec:fitting}
From the experimental results in Section~\ref{sec:results} we can see that each
LLM has a unique runtime and energy consumption characteristic that is a
function of the given workload. In this section we develop and apply these
models to optimizing energy and runtime of serving LLMs.

\subsection{Independence of Input and Output Tokens}

From observing our results, we explored whether the number of input and output
tokens are independent in their effect on the energy consumption and runtime.
The following table presents the ANOVA results for assessing the effects of the
number of input tokens, the number of output tokens, and their interaction on
the total energy consumption and runtime for LLM inference. To collect this data
we perform a grid search from 8 to 2048, in increments of powers of two, for the
space of input and output tokens to eliminate the bias of holding the input or
output size constant. This analysis includes data aggregated across all models
in Table~\ref{tab:models}.

\begin{table}[ht]
  \centering
  \caption{ANOVA Results for LLM Energy Consumption and Runtime}
  \label{tab:anova_results}
  \resizebox{\columnwidth}{!}{
    \begin{tabular}{@{}llcccc@{}}
      \toprule
      \textbf{Metric} & \textbf{Variable} & \textbf{Sum of Squares} & \textbf{F-statistic} & $p$-\textbf{value} \\ \midrule
      \multirow{3}{*}{Energy (J)} & Input Tokens & $5.17 \times 10^{10}$ & 15.86 & $3.79 \times 10^{-17}$ \\
                      & Output Tokens & $4.13 \times 10^{11}$  & 126.63 & $1.22 \times 10^{-65}$ \\
                      & Interaction & $1.18 \times 10^{11}$  & 4.53 & $4.67 \times 10^{-15}$ \\
      \midrule
      \multirow{3}{*}{Runtime (s)} &Input Tokens & $3.43 \times 10^{5}$ & 12.97 & $2.34 \times 10^{-14}$ \\
                      & Output Tokens & $2.78 \times 10^{6}$ & 104.98 & $4.56 \times 10^{-60}$ \\
                      & Interaction & $8.21 \times 10^{5}$ & 3.88 & $1.92 \times 10^{-12}$ \\ \bottomrule
    \end{tabular}}
\end{table}

The \textit{number of input tokens} and \textit{number of output tokens} both
individually have a substantial impact on energy consumption and runtime, with
output tokens having a larger effect size as indicated by the higher $F$
statistic. Also, the \textit{interaction} term shows that the input and output
tokens depend on each other while impacting energy consumption and runtime. The
high $F$-statistics and extremely low $p$-values for these effects confirm their
significance. Therefore, we conclude that there is dependence between the number
of input and output tokens for energy consumption and runtime.

\subsection{Modeling Energy and Runtime}
We use the results in Table~\ref{tab:anova_results} to guide the creation of
models to predict the energy consumption and runtime of LLMs for use in
optimization problems such as those discussed in Section~\ref{sec:routing}.

\begin{figure*}[!htb]
  \centering
  \subfigure[Energy Consumption]
  {
    \label{fig:offline-energy}
    {\includegraphics[height=31ex]{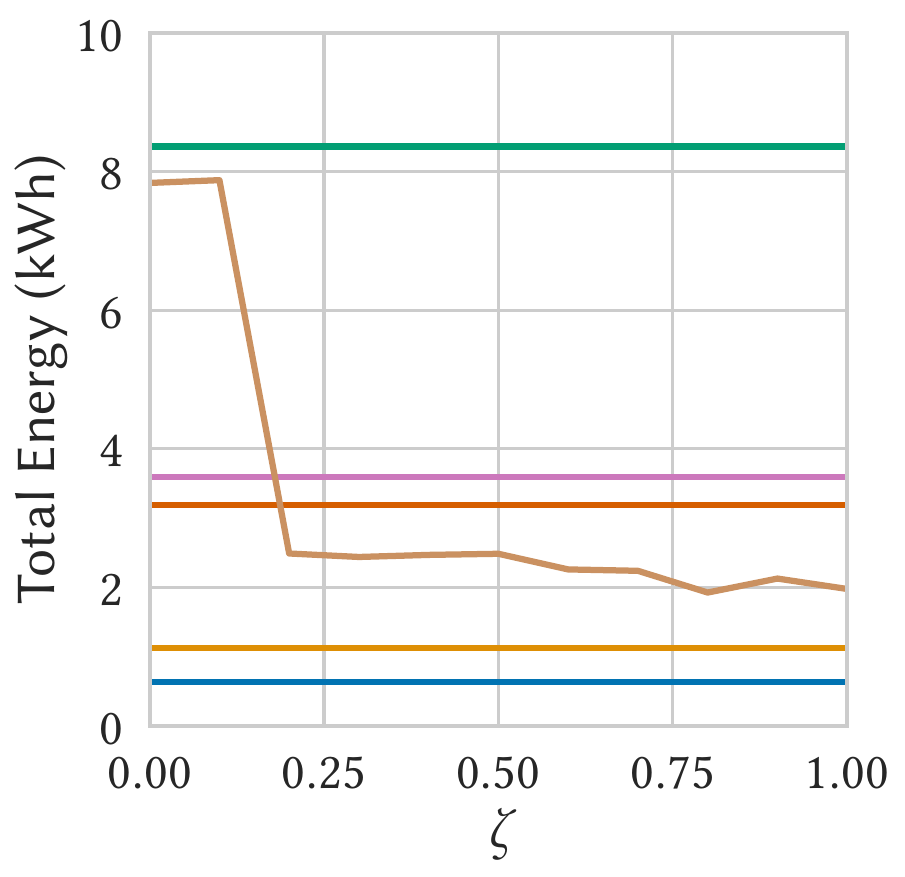}}
  }
  \subfigure[Mean Runtime]
  {
    \label{fig:offline-runtime}
    {\includegraphics[height=31ex]{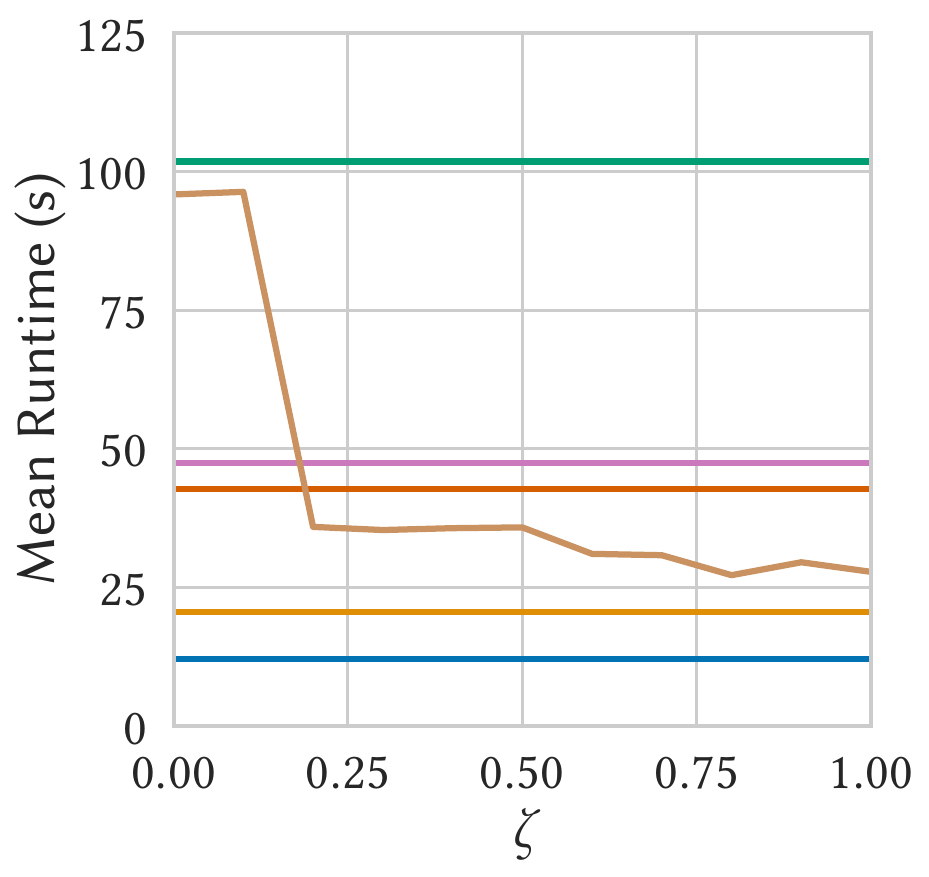}}
  }
  \subfigure[Accuracy]
  {
    \label{fig:offline-accuracy}
    {\includegraphics[height=31ex]{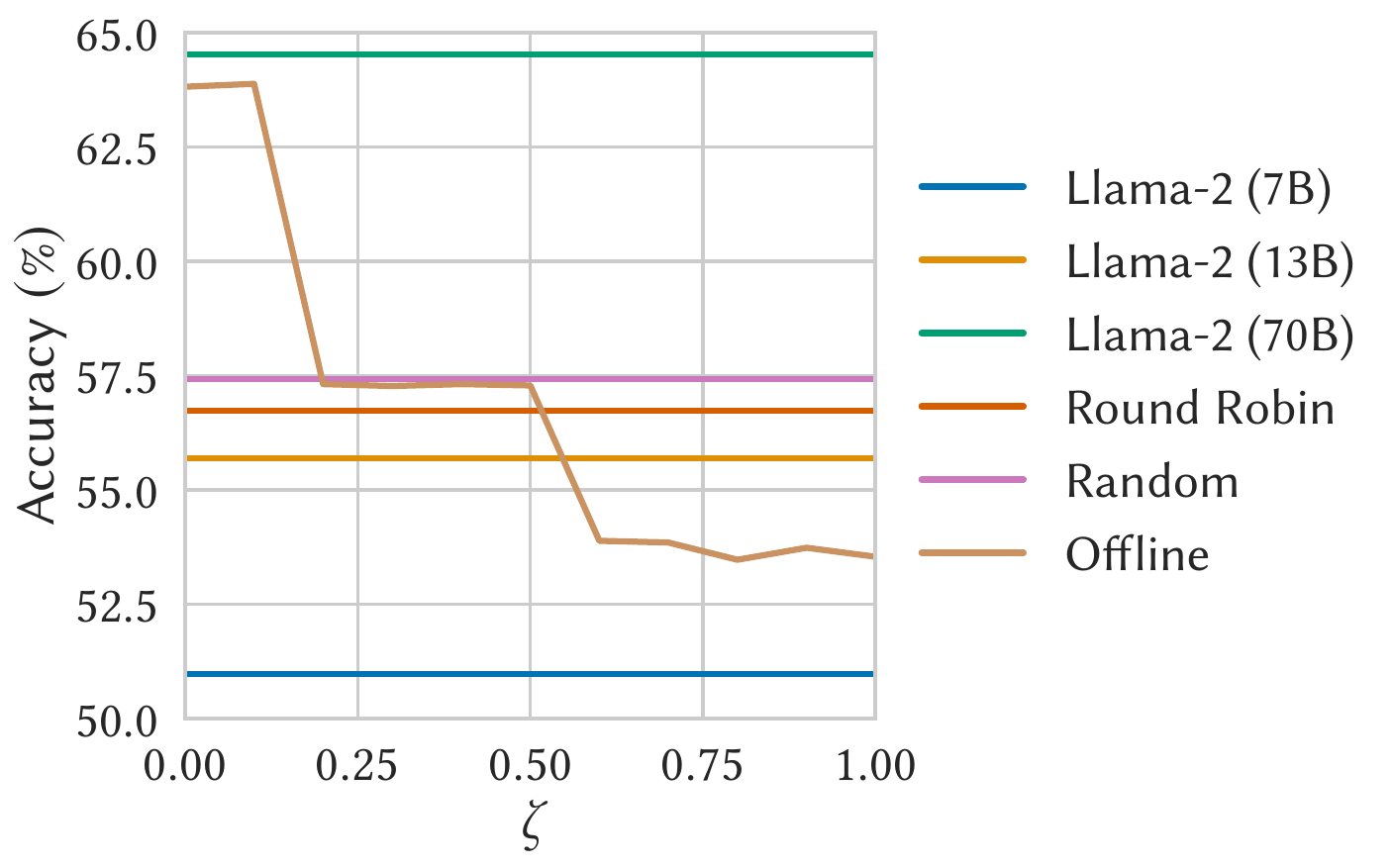}}
  }
  \caption{\label{fig:offline}Behavior under offline simulation as $\zeta$
    varies. Round-robin and Random query assignment are indistinguishable.}
\end{figure*}

For accurate models based on the number of input and output tokens there needs
to be an interaction term that combines them. We therefore propose a model to
describe the total energy consumption for a model \( K \) as a function of input
and output tokens, \( \tau_{in} \) and \( \tau_{out} \), respectively:
\begin{equation}\label{eqn:energy-model}
  e_{K}(\tau_{in}, \tau_{out}) =  \alpha_{K,0} \tau_{in} + \alpha_{K,1} \tau_{out} + \alpha_{K,2} \tau_{in} \tau_{out},
\end{equation}
where \( \alpha_{K,0}, \alpha_{K,1}, \alpha_{K,2} \) are parameters determined
through ordinary least squares (OLS) regression for each model and system
combination.

Similarly, we propose the following model to describe the total runtime for a
model \( K \) as a function of input and output tokens, \( \tau_{in} \) and
\( \tau_{out} \), respectively:
\begin{equation}\label{eqn:runtime-model}
  r_{K}(\tau_{in}, \tau_{out}) = \beta_{K,0} \tau_{in} + \beta_{K,1} \tau_{out} + \beta_{K,2} \tau_{in} \tau_{out},
\end{equation}
where \( \beta_{K,0}, \beta_{K,1}, \beta_{K,2} \) are also unique to each model
$K$.

Using the statsmodel (v0.14.2) Python package and its OLS API, we can determine
the values of $\alpha_{K,i}$ and $\beta_{K,j}$ that best fit
Equations~\ref{eqn:energy-model}~and~\ref{eqn:runtime-model} for each LLM, $K$.
A summary of the quality of these fits are included in Table~\ref{tab:fit}. As
we can see, this model has high explainability for the effect of input and
output tokens on energy and runtime for inference of these different LLMs.


\begin{table}[ht]
  \centering
  \caption{Summary of OLS Regression Results Across Models}
  \label{tab:fit}
  \resizebox{\columnwidth}{!}{
    \begin{tabular}{lrrrrrrr}
      \toprule
      \textbf{LLM (\# Params)} & \multicolumn{3}{c}{\textbf{Energy Model} ($e_K$)} & \multicolumn{3}{c}{\textbf{Runtime Model} ($r_K$)} \\
      \cmidrule(lr){2-4} \cmidrule(lr){5-8}
                               & \boldmath$R^2$ & \textbf{F-statistic} & $p$\textbf{-value} & \boldmath$R^2$ & \textbf{F-statistic} & $p$\textbf{-value} &\\
      \midrule
      Falcon (7B)     & 0.964  & 681.2  & 2.53e-55  & 0.962  &  651.1 & 1.35e-54 \\
      Falcon (40B)    & 0.972  & 904.5  & 1.78e-60  & 0.976  & 1073.0 & 2.74e-63 \\
      Llama-2 (7B)    & 0.973  & 942.3  & 3.76e-61  & 0.972  & 1032.0 & 1.19e-62 \\
      Llama-2 (13B)   & 0.972  & 887.8  & 3.60e-60  & 0.972  &  907.0 & 1.60e-60 \\
      Llama-2 (70B)   & 0.976  & 1022.0 & 6.66e-62  & 0.980  & 1230.0 & 6.23e-65 \\
      Mistral (7B)    & 0.975  & 997.0  & 1.70e-61  & 0.976  & 1039.0 & 3.62e-62 \\
      Mixtral (8x7B)  & 0.980  & 1238.0 & 4.97e-65  & 0.992  & 3139.0 & 2.23e-80 \\
      \bottomrule
    \end{tabular}}
\end{table}

\subsection{Applying Our Models to Workload Routing}

We can now use our runtime and energy consumption models to solve the
workload-aware routing problem outlined in Section~\ref{sec:routing}. Using PuLP
(v.2.8.0), a Python package designed for solving optimization problems like that
we formulate in Equation~\ref{eqn:opt-2-1}, we can encode a workload of input
and output tokens with a set of binary variables that indicate which model will
process that pair of tokens. Then, we convert the given constraints in
Equations~\ref{eqn:opt-2-2}--\ref{eqn:opt-2-4} using this format and effectively
route our workload to different models.

As we show in Table~\ref{tab:models} and Figures~\ref{fig:input-token-results}
and~\ref{fig:output-token-results}, an LLM with a larger parameter count has
greater accuracy but also greater runtime and energy consumption for each input
and output token. It is reasonable to host differently sized models to allow us
to serve inference requests more runtime and energy efficiently with a trade-off
of slightly lower accuracy.

For this example, we consider a data center serving the three Llama-2 models of
7B, 13B, and 70B parameters. Assume that our set $\mathcal{K} = \{1, 2, 3\}$
enumerates those models, respectively. A tunable parameter that affects our
optimization problem is the data center partition $\gamma_i.$ In our evaluation,
we choose $\gamma_1 = 0.05, \gamma_2 = 0.2,$ and $\gamma_3 = 0.75.$

With this, we can use the model for energy consumption of each LLM, $K$, in
Equation~\ref{eqn:energy-model} and our function to capture accuracy from
Equation~\ref{eqn:accuracy} to calculate the costs associated with each query
and model as shown in Equation~\ref{eqn:opt-2-1}. For our sample workload, we
use a subset of 500 queries from the Alpaca dataset~\cite{alpaca}, as it is a
collection of 52002 queries with answers from GPT-4~\cite{openai2023gpt4}.

Figure~\ref{fig:offline} shows the trade-offs in energy consumption, runtime,
and accuracy by varying the operational parameter $\zeta$ while routing queries
to different models. We represent as constants (straight lines) methods that do
not use $\zeta$, preferring to pick either a single LLM or to use a simple
query-independent mechanism to route a query to an LLM. The remaining
non-constant line represents the trade-off our offline scheduler makes as it
adjusts to changes in $\zeta$.


In Figure~\ref{fig:offline-energy}, we see that energy consumption is high when
$\zeta$ is low because the system prioritizes accuracy over energy efficiency.
Higher $\zeta$ values lead to more energy-efficient routing decisions,
sacrificing accuracy for energy savings. Similarly,
Figure~\ref{fig:offline-runtime} shows that the mean runtime per query decreases
with increasing $\zeta$. A low $\zeta$ value results in longer runtimes as the
system routes queries to models that provide higher accuracy but are less
efficient in time and energy. Conversely, higher $\zeta$ values result in
shorter runtimes, as the system favors more energy and time-efficient models
over the most accurate ones. Figure~\ref{fig:offline-accuracy} demonstrates the
accuracy-cost trade-off, with small increases in accuracy requiring significant
increases in runtime and energy consumption.

Our solution allows data center operators use $\zeta$ to navigate the trade-off
space by, e.g.,~providing higher accuracy when energy prices are lower, or
delivering lower latency and lower energy responses during times of peak load
albeit with slightly reduced accuracy. This flexibility is important for
adapting to different operational scenarios.

\section{Conclusions}\label{sec:conclusions}

In this paper, we have examined the significant energy expenditure of LLM
inference. We show that modeling and optimizing the energy consumption of LLM
inference for a system is straightforward. We also showed that SMoE LLMs exhibit
very promising energy efficiency characteristics. Through our models of energy
and runtime we contribute to the ongoing efforts towards sustainable AI by
providing a tunable optimization framework that allows for system operators to
trade-off energy and accuracy. We confirm our hypothesis that there is potential
for energy optimization using models of energy and accuracy.

Of course, as many others have done~\cite{lin2024exploding, chien2023reducing,
  lin2023adapting, henderson2020towards,
  satveer2016comparative, li2024sustainable} we have used energy consumption as a proxy for carbon
footprint. As pointed out by Kannan and Kremer~\cite{kannan2023towards},
improving carbon efficiency and energy efficiency are distinct goals, yet they
are related and energy metrics can assist in understanding the magnitude of
emissions for a given application~\cite{anderson2023treehouse}. Our measurements
of energy consumption are also based on a single node in an HPC setting and so
we cannot capture the runtime and energy overheads introduced by faults,
networking, and communications that would pertain at data center scale. We also
disabled key-value caching~\cite{pope2022efficiently} to establish a performance
baseline; future work should explore the impact of this and other optimizations.
Finally, our workload-models are specific primarily to an NVIDIA A100 (40GB), as
pointed out in other studies there are large variations for the same inference
task across hardware~\cite{samsi2023words,wilkins2024hybrid}.

We hope that our energy models can be used in real-time systems to reduce energy
consumption dynamically. By integrating these models into online scheduling
algorithms, data centers can make energy-aware decisions based on the current
workload and system state. This real-time optimization approach has the
potential to significantly improve the energy efficiency of LLM inference in
production environments. Similarly, including externalities like energy pricing
and availability of sustainable energy into our model would bring systems closer
to meeting sustainability goals.

\begin{acks}
  We gratefully acknowledge the computing resources provided on Swing, a
  high-performance computing cluster operated by the Laboratory Computing
  Resource Center at Argonne National Laboratory. During this work GW was
  supported by a Churchill Scholarship. We would like to thank the reviewers for
  their valuable feedback to help improve our work.
\end{acks}

\bibliographystyle{ACM-Reference-Format}
\bibliography{hotcarbon}

\end{document}